\PassOptionsToPackage{sort&compress}{natbib}  
\documentclass[12pt]{elsarticle}
\usepackage{booktabs}
\usepackage{caption}

\usepackage{amsmath}
\usepackage{bm}
\usepackage[usenames, dvipsnames]{xcolor}
\usepackage[utf8x]{inputenc}
\usepackage[T1]{fontenc}
\usepackage{tikz}
\usepackage{pgfplots, pgfplotstable}
\usetikzlibrary{calc,arrows,fadings,decorations.pathreplacing,decorations.markings,patterns,shapes.geometric}
\usepackage{fp}
\usepackage{yhmath}
\usepackage{verbatim}
\usepackage{svg}
\usepackage{amssymb}
\usepackage{caption}
\usepackage{subcaption}
\usetikzlibrary{calc,fit,intersections,shapes}
\usetikzlibrary{calc}
\usepgfplotslibrary{polar}
\pgfplotsset{compat=1.6}
\pgfplotsset{every axis plot}
\usepackage{amsmath}
\usepackage{setspace}
\usepackage{lineno}

\usepackage{algorithm}
\usepackage{algpseudocode}

\newcommand{\bfM}{\mathbf{M}}
\newcommand{\bfK}{\mathbf{K}}
\newcommand{\bfC}{\mathbf{C}}
\newcommand{\bfF}{\mathbf{F}}
\newcommand{\bfx}{\mathbf{x}}
\newcommand{\bfR}{\mathbf{R}}
\newcommand{\bfT}{\mathbf{T}}
\newcommand{\bfI}{\mathbf{I}}
\newcommand{\bfth}{\bm{\theta}}
\newcommand{\bfTh}{\bm{\Theta}}
\newcommand{\bfPhi}{\bm{\Phi}}
\newcommand{\bfphi}{\bm{\phi}}
\newcommand{\bfeta}{\bm{\eta}}
\newcommand{\bfPsi}{\bm{\Psi}}

\newcommand{\RPcondition}{(\bfR^o)^{T}\bfPhi^p}
\newcommand{\bfQ}{\mathbf{Q}}
\newcommand{\bfX}{\mathbf{X}}
\newcommand{\bfY}{\mathbf{Y}}

\begin{document}

\title{Multi-Region Matrix Interpolation for Dynamic Analysis of Aperiodic Structures under Large Model Parameter Perturbations}

\author[label1]{J. Pereira}
\author[label1]{R. O. Ruiz \footnote{Corresponding author}}

\address[label1]{Department of Mechanical Engineering, University of Michigan-Dearborn, Dearborn, USA}

\begin{singlespace}
\begin{abstract}

This work introduces a surrogate-based model for efficiently estimating the frequency response of dynamic mechanical metamaterials, particularly when dealing with large parametric perturbations and aperiodic substructures. The research builds upon a previous matrix interpolation method applied on top of a Craig-Bampton modal reduction, allowing the variations of geometrical features without the need to remesh and recompute Finite Element matrices. This existing procedure has significant limitations since it requires a common modal projection, which inherently restricts the allowable perturbation size of the model parameters, thereby limiting the model parameter space where matrices can be effectively interpolated. The present work offers three contributions: (1)  It provides structural dynamic insight into the restrictions imposed by the common modal projection, demonstrating that ill-conditioning can be controlled, (2) it proposes an efficient, sampling-based procedure to identify the non-regular boundaries of the usable region in the model parameter space, and (3)  it enhances the surrogate model to accommodate larger model parameter perturbations by proposing a multi-region interpolation strategy. The efficacy of this proposed framework is verified through two illustrative examples. The first example, involving a unit cell with a square plate and circular core, validates the approach for a single well-conditioned projection region. The second example, using a beam-like structure with vibration attenuation bands, demonstrates the true advantage of the multi-region approach, where predictions from traditional Lagrange interpolation deviated significantly with increasing perturbations, while the proposed method maintained high accuracy across different perturbation levels.

\end{abstract}
\end{singlespace}



\maketitle

\section{Introduction}
\label{S:1}

Dynamic mechanical metamaterials have garnered significant research interest due to their unique capabilities in controlling acoustic and vibration transmission, as highlighted by Wu et al. \cite{wu2021brief}. These materials are constructed from the spatial assembly of building blocks or substructures, whose meticulously designed geometries and selected materials enable precise manipulation of mechanical energy. When the building blocks are identical, the analysis of these materials can be performed at the substructure level by applying periodic boundary conditions based on the Bloch theorem \cite{phanidynamics}. However, for nearly-periodic or aperiodic arrangements of substructures (e.g., graded metamaterials \cite{jian2023analytical}, waveguides \cite{li2019active}, or metamaterials under manufacturing uncertainties \cite{cantero2022robust}), the performance analysis must be conducted at the structure level (or full-scale modeling). Such configurations enable individual substructures to possess different geometrical features and material properties, thereby capturing the effect of between-substructure variability on overall performance.  However, this often leads to a substantial increase in computational load, as each substructure must be individually meshed and assembled. Consequently, this complexity makes the recurrent analysis of metamaterial structures using high-fidelity simulations (e.g., in topological optimization \cite{watts2019simple} and uncertainty quantification \cite{henneberg2020periodically}) difficult due to the increased number of substructure geometries that need to be tested \cite{sinha2023programmable}. To mitigate these challenges, researchers are increasingly focused on developing efficient surrogate models that can predict the dynamic performance of metamaterials at the structural level with reduced computational costs, as discussed by Cerniauskas et al. \cite{cerniauskas2024machine}.

\begin{figure}[ht]h
    \centering
    \includegraphics[width=1\linewidth]{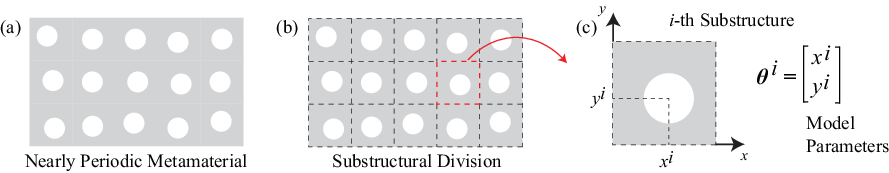}
    \caption{a) Example metamaterial structure and b) illustration of the substructure model parameter vector $\bfth^i$ and its components $x^i$ and $y^i$.}
    \label{F1}
\end{figure}

\subsection{Matrix Interpolation Surrogate Models}
Among the available surrogate models, the approach presented by Mencik in \cite{mencik2021model} and further elaborated in \cite{mencik2024improved} is particularly relevant for designing dynamic mechanical metamaterials since it enables the exploration of nearly periodic metamaterials at a low computational cost. It introduces a Finite Element (FE)-based technique that permits the variation of geometrical features at the substructural level without explicit remeshing or recomputing their high-fidelity FE matrices. Essentially, this method approximates the mass and stiffness matrices at the substructure level when geometrical variations are introduced. Subsequently, these estimated matrices can be leveraged (after proper assembly) to approximate the Frequency Response Function (FRF) of a full-scale structure. The procedure involves dividing the structure into its constituent substructures, each parameterized by a set of geometrical features or model parameters (e.g., length, width, or thickness) as depicted previously in Figure~\ref{F1}. At the substructure level, a modal reduction of the FE matrices is performed using the Craig-Bampton (CB) method \cite{craig1968coupling}, yielding reduced mass and stiffness matrices. These reduced FE matrices are computed at $P$ predefined set of model parameters, also known as support points and denoted here as $\{\bfth^p;p=1,...,P\}$, which are then used to estimate the matrices for any new set of model parameters $\bfth$ through matrix interpolation.

\begin{figure}[ht]
    \centering
    \includegraphics[width=0.9\linewidth]{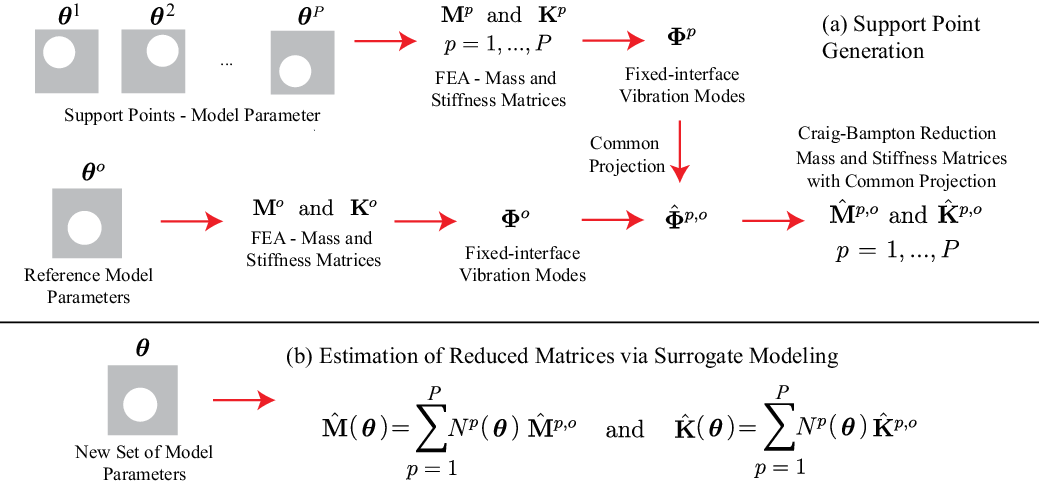}
    \caption{Schematic representation of the surrogate model presented in \cite{mencik2024improved} for predicting substructural Craig-Bampton matrices.}
    \label{F2}
\end{figure}

The procedure introduced by Mencik in \cite{mencik2021model,mencik2024improved} can be schematized as presented in Figure~\ref{F2}. It comprises two distinctive blocks: the generation of support points (Figure~\ref{F2}(a)), and the use of support points to approximate the FE matrices for new sets of model parameters (Figure~\ref{F2}(b)). In the first block, these sets of support points require high-fidelity analysis using standard FE, which leads to sets of mass and stiffness matrices, $\{{\bfM}^p; p=1,...,P\}$ and $\{{\bfK}^p; p=1,...,P\}$, respectively. Following the CB method, a subset of vibration modes at fixed interface is computed and retained for each support point, leading to $\{\bfPhi^p; p=1,...,P\}$. This methodology has a particularity: in addition, it requires the use of a reference set of model parameters within the support points, denoted here as ${\bfth}^o$, as well as its respective subset of vibration modes at fixed interface $\bfPhi^o$. This reference is used to project the vibration modes of each support point ($\bfPhi^p$) over a common modal basis. Ultimately, the modal projection of ${\bfM}^p$ and ${\bfK}^p$ is performed over a basis denoted as $\hat{\bfPhi}^{p,o}$ to explicitly indicate that it corresponds to the projection of substructure $p$ over the reference $o$, leading to $\{\hat{\bfM}^{p,o}; p=1,...,P\}$ and $\{\hat{\bfK}^{p,o}; p=1,...,P\}$. A detailed description of this reduction is offered later in Sections~\ref{S:2.1} and ~\ref{S:2.2}. In the second block, the reduced mass and stiffness matrices $\hat{\bfM}$ and $\hat{\bfK}$ for a new set of model parameters $\bfth$ can be estimated by interpolating the precomputed matrices $\hat{\bfM}^{p,o}$ and $\hat{\bfK}^{p,o}$ using, in this example, a two-dimensional interpolation scheme with interpolation functions $N^p$. 

Despite its proven effectiveness in estimating metamaterial FRFs, this procedure exhibits certain limitations, also acknowledged in \cite{mencik2024improved}. Firstly, the substructure's FE mesh must be parameterized with respect to the geometry, meaning that nodes and elements must be distorted while retaining their original connectivity. Secondly, to enable interpolation, the CB matrices (of any sub-structure) require projection onto a common modal space, i.e., $\{\hat{\bfM}^{p,o}; p=1,...,P\}$ and $\{\hat{\bfK}^{p,o}; p=1,...,P\}$. This common modal projection inherently restricts the allowable perturbation size of the model parameters, $\Delta\bfth^p=|\bfth^p-\bfth^o|$, thereby limiting the model parameter space where the CB matrices can be effectively interpolated. Crucially, \cite{mencik2021model,mencik2024improved} offers no explicit strategies to identify the maximum permissible perturbation for valid interpolation—i.e., the boundary in the model parameter space that defines the usable region for the surrogate model. This raises the need to not only identify the interpolative space but also to extend this common modal projection technique to accommodate larger model parameter perturbations.

\subsection{Contributions}
The present work builds upon the procedure outlined in \cite{mencik2021model,mencik2024improved} and offers three primary contributions: (1) it provides structural dynamic insight into the restrictions imposed by the common modal projection; (2) it proposes a procedure to identify the boundaries of the usable region in the model parameter space where the surrogate model can be reliably deployed; and (3) it enhances the surrogate model to accommodate larger model parameter perturbations by proposing a sample-based procedure to identify regions in the model parameter space where surrogate models can be deployed. These three aspects are addressed in the following sections and represent a significant step forward in understanding and improving these interpolation-based surrogate models.

The paper is organized as follows. Section 2 provides a general discussion of the common modal projection, including its restrictions and characterization. In Section 3, a procedure is proposed for identifying the boundary of the region where the surrogate model can be deployed using a sample-based model parameter space exploration algorithm. Section 4 then proposes a procedure to accommodate larger model parameter spaces for deploying surrogate models. Section 5 presents an illustrative example involving 2D plates with geometrical model parameters extracted from \cite{mencik2024improved} for verifying and comparing the proposed techniques. Subsequently, Section 6 presents a second illustrative example using 2D shells to demonstrate the enhanced model parameter space achieved with the proposed framework. Finally, Section 7 concludes the paper and discusses potential future research opportunities.

\section{Ill-conditioned Modal Projections observed in the Substructuring Approach}
\label{S:2}

\subsection{The Craig-Bampton Reduction}
\label{S:2.1}

The general structure of the CB method \cite{craig1968coupling} adopted to generate the support points described in the first block of Figure~\ref{F2} is established as follows. Let's consider the $p$-th substructure parameterized by model parameters $\bfth^p$. After applying a standard FEA procedure, the equation of motion in the absence of damping can be expressed as in Eq.~(\ref{eq1}), with $\bfM^p$ and $\bfK^p$ representing the mass and stiffness matrices, respectively, while $\bfx^p$ corresponds to the vector of degrees of freedom (DoF), $\ddot{\bfx}^p$ its second time derivative, and $\bfF^p$ the forcing vector. This representation can be extended as presented in Eq.~(\ref{eq2}), where subscripts $i$ and $j$ correspond to interface and non-interface DoF, respectively. In this context, the interface is understood as the DoF shared by two adjacent substructures.
\begin{equation}
    \bfM^p\ddot{\bfx}^p + \bfK^p\bfx^p=\bfF^p.
    \label{eq1}
\end{equation}
\begin{equation}
    \begin{bmatrix}  
        \bfM_{ii}^p & \bfM_{ij}^p \\
        \bfM_{ji}^p & \bfM_{jj}^p
    \end{bmatrix}
    \begin{bmatrix}
        \ddot\bfx_i^p \\
        \ddot\bfx_j^p
    \end{bmatrix} +
    \begin{bmatrix}  
        \bfK_{ii}^p & \bfK_{ij}^p \\
        \bfK_{ji}^p & \bfK_{jj}^p
    \end{bmatrix}    
    \begin{bmatrix}
        \bfx_i^p \\
        \bfx_j^p
    \end{bmatrix} =
    \begin{bmatrix}
        \bfF_i^p \\
        \bfF_j^p
    \end{bmatrix}.
    \label{eq2}
\end{equation}
\vspace{1pt}

The basis of the CB method corresponds to the projection of the physical coordinates $\bfx^p$ onto a mixture of: (1) physical coordinates representing the DoF at the substructure's interface  $\bfx^p_i$, and (2) modal coordinates $\bfeta^p$ obtained by projecting the non-interface DoF $\bfx^p_j$ over a subset of vibration modes at fixed interface $\bfPhi^p$. This projection is established by the transformation presented next in Eq.~(\ref{eq3}):
\begin{equation}
    \begin{bmatrix}
        \bfx_i^p \\
        \bfx_j^p
    \end{bmatrix} = 
    \begin{bmatrix}
        \bfI & \mathbf{0} \\
        \bfPsi^p & \bfPhi^p
    \end{bmatrix}
    \begin{bmatrix}
        \bfx_i^p \\
        \bfeta^p
    \end{bmatrix} \Rightarrow \bfx^p = \bfT^p \hat{\bfx}^p,
    \label{eq3}
\end{equation}
\vspace{1pt}

\noindent where $\bfPsi^p$ is defined following Eq.~(\ref{eq4}) (capturing the static interaction between interface and internal DoF), while $\bfPhi^p$ is defined by the generalized eigenvalue problem presented in Eq.~(\ref{eq5}). Here, $\bfphi^p$ corresponds to the vibration mode, $\lambda^p$ to its corresponding eigenvalue, the subscript $l$ refers to the mode number, $q$ indicates the total number of fixed-interface modes retained, and $n$ corresponds to the total number of non-interface DoF. Notice that $q$ can be lower than $n$, which corresponds to a modal reduction.
\begin{equation}
    \bfPsi^p=-\text{inv}(\bfK_{jj}^p)\bfK_{ji}^p.
    \label{eq4}
\end{equation}
\begin{equation}
    \begin{array}{cc}
        (\bfM_{jj}^p-\lambda_l^p\bfK_{jj}^p)\bfphi_l^p=\mathbf{0},\\
        \bfPhi^p=[\bfphi_1^p, ... ,\bfphi_q^p],\ q\leq n.
    \end{array}
    \label{eq5}
\end{equation}
\vspace{1pt}

After introducing the transformation presented in Eq.~(\ref{eq3}) into the equation of motion (Eq.~\ref{eq1}), the reduced system of equations takes the following form:
\begin{equation}
    \hat{\bfM}^p
    \ddot{\hat{\bfx}}^p +
    \hat{\bfK}^p
    \hat{\bfx}^p =
    \hat{\bfF}^p,
    \label{eq6}
\end{equation}

\noindent where $\hat{\bfM}^p$ and $\hat{\bfK}^p$ correspond to the reduced mass and stiffness matrices, respectively, while $\hat{\bfF}^p$ refers to the reduced forcing vector, all of them obtained as:
\begin{equation}
    \begin{array}{cc}
        \hat{\bfM}^p = (\bfT^p)^T\bfM^p\bfT^p,\\
        \hat{\bfK}^p = (\bfT^p)^T\bfK^p\bfT^p,\\
        \hat{\bfF}^p = (\bfT^p)^T\bfF^p,
    \end{array}
    \label{eq7}
\end{equation}

\noindent where superscript $T$ indicates matrix transpose.
 
\subsection{Common Modal Basis Transformation}
\label{S:2.2}

The matrices presented in Eq.~(\ref{eq7}) and obtained using the standard CB method \cite{craig1968coupling} cannot be directly interpolated because they are not projected onto a common modal basis. Thus, Mencik \cite{mencik2021model} proposed a modification to the standard CB transformation $\bfT^p$ (Eq. \ref{eq3}) inspired by the work of Panzer et al. \cite{PanzerMohringEidLohmann+2010+475+484}. The modification consists of avoiding the use of $\bfPhi^p$ to compute the reduced mass and stiffness matrices. Instead, it is proposed to incorporate a reference substructure (previously highlighted in Figure~\ref{F2} as $\bfth^o$) to project the fixed-interface vibration modes $\bfPhi^p$ over a common basis $\bfR^o$ defined as:
\begin{equation}
    \bfR^o=\bfM_{jj}^o\bfPhi^o.
    \label{eq8}
\end{equation}

The common basis $\bfR^o$ is normalized by $\bfM_{jj}^o$ following the procedure presented in \cite{mencik2024improved}. However, other normalizations can be implemented as long as orthogonality is met, i.e., $(\bfR^o)^T \bfR^o=\bfI$; see \cite{mencik2021model} for a more detailed discussion. As a consequence, the fixed-interface vibration modes of the $p$-th substructure projected over the  common basis take the following form:
\begin{equation}
    \hat{\bfPhi}^{p,o} = \bfPhi^p[(\bfR^o)^{T}\bfPhi^p]^{-1}.
    \label{eq9}
\end{equation}

Note that the superscript $p,o$ is introduced here to explicitly indicate the dependency of $\hat{\bfPhi}^{p,o}$ on $\bfth^p$ and $\bfth^o$. This notation will play a significant role in subsequent sections. With this common modal basis, the CB transformation matrix from Eq.~(\ref{eq3}) is modified as follows:
\begin{equation}
    \hat{\bfT}^{p,o} = 
    \begin{bmatrix}
        \bfI & \mathbf{0} \\
        \bfPsi^p & \hat{\bfPhi}^{p,o}
    \end{bmatrix},
    \label{eq10}
\end{equation}

\noindent leading to the following reduced forcing vector, and mass and stiffness matrices (a detailed derivation of this transformation matrix can be found in \cite{mencik2021model}): 
\begin{equation}
    \begin{array}{cc}
        \hat{\bfM}^{p,o} = (\bfT^{p,o})^T\bfM^p\bfT^{p,o},\\
        \hat{\bfK}^{p,o} = (\bfT^{p,o})^T\bfK^p\bfT^{p,o},\\
        \hat{\bfF}^{p,o} = (\bfT^{p,o})^T\bfF^p.
    \end{array}
    \label{eq11}
\end{equation}

The reduced vector and matrices presented in Eq.~(\ref{eq11}) are then used in the interpolation scheme presented in the second block of Figure~\ref{F2}. However, for Eq.~(\ref{eq9}) to be valid, the matrix product $\RPcondition$ must be invertible, which is the primary constraint of these common basis projections. As discussed in \cite{mencik2021model}, this requirement implies that the column ranges of the matrices $(\bfR^o)^T$ and $\bfPhi^p$ are sufficiently close, often translating to small mesh distortions. However, this mathematical constraint does not fully explain the physical significance of having "sufficiently close" column ranges. As $\bfth^p$ progressively deviates from $\bfth^o$, these modes will change, gradually differing from the nominal modes. At a particular model parameter perturbation $\Delta\bfth_{max}^p$, phenomena such as mode crossing, veering, and coalescence may occur between retained and neglected modes in Eq.~(\ref{eq5}) \cite{gallina2011enhanced}. These phenomena introduce new information about the fixed-interface dynamics, offering potential causes of the deviations between the modal information contained in $\bfR^o$ and $\bfPhi^p$. The following section tackles explicitly the effect of mode crossing on the numerical stability of the CB method under a common modal basis. 

\subsection{Mode Crossing Effect on Modal Projection Conditioning}
\label{S:2.3}

\begin{figure}[ht]
    \centering
    \includegraphics[width=0.8\linewidth]{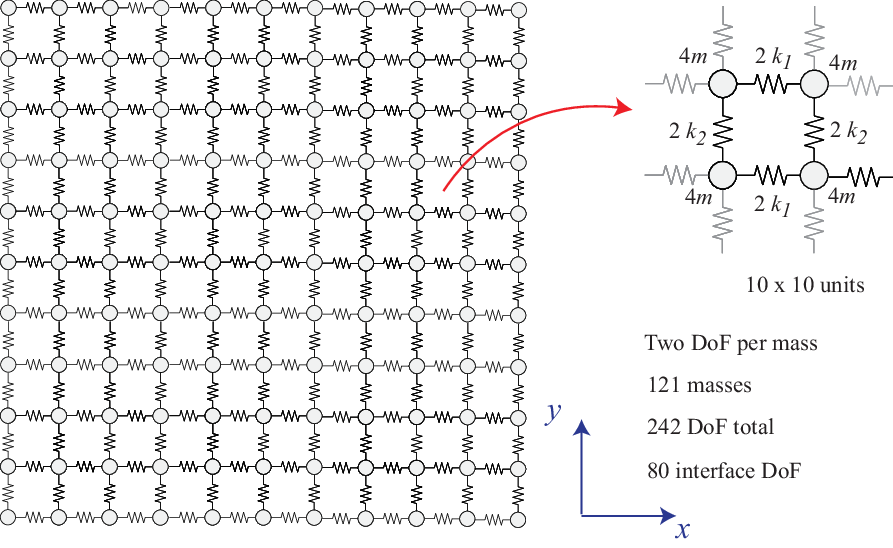}
    \caption{Substructure used to study the effect of mode switching on the ill-conditioning of $\hat{\bfPhi^{p,o}}$.}
    \label{F3}
\end{figure}

The mode crossing and its influence on the invertibility of $\RPcondition$ is studied here by using a simplified two-dimensional substructure, consisting of multiple mass-spring systems. The chosen substructure, depicted in Figure~\ref{F3}, comprises 121 masses and 242 DoF. All horizontal springs within the system are identical, as are all vertical springs, although their stiffness values differ from those of the horizontal springs. This substructure was specifically selected for its large number of DoF, which provides ample scope to explore the impact of varying the number of retained modes, denoted as $q$ in Eq.~(\ref{eq5}). Two distinct cases are analyzed: Case 1 demonstrates an ill-conditioning behavior that is independent of the perturbation size (defined as the distance between the support point $\bfPhi^p$ and the reference $\bfPhi^o$), but is directly dependent on $q$; whereas Case 2 illustrates an ill-conditioning that arises from an increment in the perturbation size for a fixed number of retained modes $q$.

\begin{figure}[ht]
    \centering
    \includegraphics{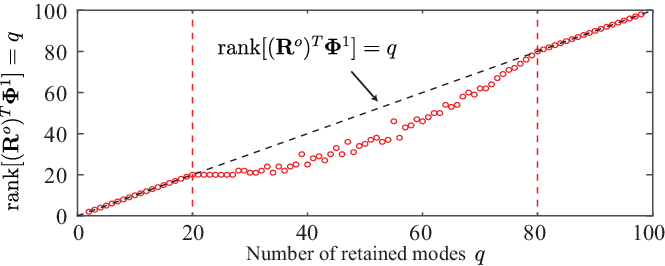}
    \caption{Rank of $(\bfR^o)^T\bfPhi^1$ for different numbers of retained modes while randomly switching all vibration modes of $\bfPhi^1$ above the $21^{\text{st}}$ and below the $80^{\text{th}}$.}
    \label{F4}
\end{figure}

\textit{Case 1}. In this case, the study is conducted over a single support point $\bfth^1$ that will be forced to match the reference configuration $\bfth^o$, i.e., imposing a zero perturbation. The model parameters in this case corresponds to $\bfth^1=\bfth^o=[m,k_1,k_2]$, with \mbox{$m=5$ g}, \mbox{$k_1=1$ kN/mm}, and \mbox{$k_2=0.9$ kN/mm}. $(\bfR^o)^T$ and $\bfPhi^1$ are computed for different numbers of retained fixed-interface modes $q$, while the columns of $\bfPhi^1$ are randomly switched between the $21^{\text{st}}$ and $80^{\text{th}}$ modes before modal truncation. The rank of $(\bfR^o)^T\bfPhi^1$ is then tracked as $q$ increases. Here, the rank indicates whether the support point can be projected into the common basis or not, in other words, a full rank assures that $\hat{\bfPhi}^{p,o}$ is well-conditioned. The results for the rank of $(\bfR^o)^T\bfPhi^1$ are depicted in Figure~\ref{F4}, where $(\bfR^o)^T\bfPhi^1$ maintains full rank (i.e., $\text{rank}[(\bfR^o)^T\bfPhi^1]=q$) for $q<21$ and $q>79$. In contrast, for $21<q<79$, $(\bfR^o)^T\bfPhi^1$ is rank deficient, suggesting a relationship between mode swapping and the ill-conditioning of $(\bfR^o)^T\bfPhi^1$. It is important to note that mode crossing between retained modes does not affect the conditioning of $(\bfR^o)^T\bfPhi^1$, i.e., the rank is full despite having mode swapping within the retained modes.

\begin{figure}[ht]
    \centering
    \includegraphics{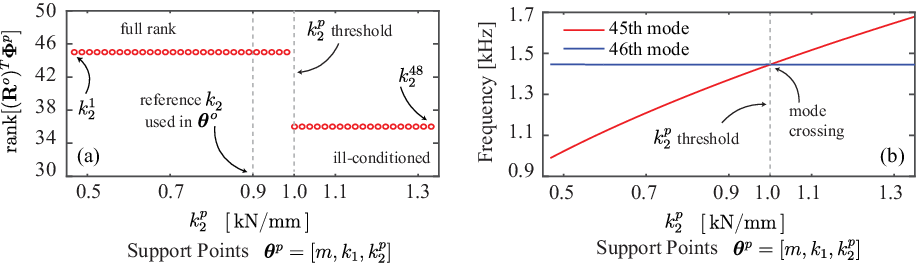}
    \caption{(a) Rank of $(\bfR^o)^T\bfPhi^p$ for each support point $k_2^p$. (b) values of the $45^{\text{th}}$ and $46^{\text{th}}$ natural frequencies each support point. Values adopted for $\bfth^o$ corresponds to: $m=5$ g, $k_1=1$ kN/mm, and $k_2=0.9$ kN/mm.}
    \label{F5}
\end{figure}

\begin{figure}[ht]
    \centering
    \includegraphics{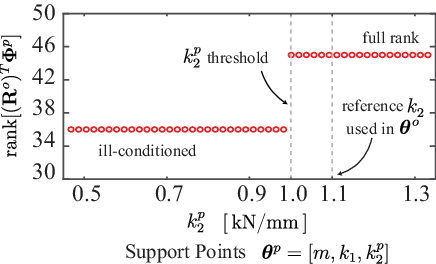}
    \caption{Rank of $(\bfR^o)^T\bfPhi^p$ as the value of $k_2^p$ is perturbed. Values adopted for $\bfth^o$ corresponds to: $m=5$ g, $k_1=1$ kN/mm, and $k_2=1.1$ kN/mm.}
    \label{F6}
\end{figure}

\textit{Case 2}. In this scenario, the reference configuration remains identical to Case 1, defined by $\bfth^o=[m,k_1,k_2]$, where $m=5$ g, $k_1=1$ kN/mm, and $k_2=0.9$ kN/mm. In contrast to Case 1, the analysis in Case 2 involves a total of 48 support points, represented as $\{\bfth^p=[m,k_1,k_2^p];\,p=1,...,48\}$. For these support points, $m$ and $k_1$ retain their respective reference values, while $k_2^p$ is perturbed within an interval of $k_2^p\in[0.5k_2,1.5k_2]$. Thus, the support points effectively correspond to perturbations introduced solely in the $k_2$ parameter around the reference configuration. A key difference from Case 1 is that now the number of retained vibration modes is fixed at $q=45$; this selection is arbitrary, serving primarily to illustrate the mode crossing phenomenon. For each support point $p$, the $\text{rank}[(\bfR^o)^T\bfPhi^p]$ is computed and presented in Figure~\ref{F5}(a). Additionally, Figure~\ref{F5}(b) displays the $45^{\text{th}}$ and $46^{\text{th}}$ fixed-interface natural frequencies of the substructures used as support points, plotted against $k_2^p$. First, it is necessary to notice that a well-conditioned projection implies $\text{rank}[(\bfR^o)^T\bfPhi^p]=45$ since it must match the number of retained modes employed. From Figure~\ref{F5}(a), it is clear that $(\bfR^o)^T\bfPhi^p$ is well-conditioned for $k_2^p<1.0$~kN/mm, becoming rank deficient beyond that threshold. Simultaneously, Figure~\ref{F5}(b) shows modes $45^{\text{th}}$ and $46^{\text{th}}$ crossing at $k_2^p=1.0$~kN/mm, indicating that this mode crossing leads to $(\bfR^o)^T\bfPhi^p$ becoming ill-conditioned. This result identifies the set of $k_2^p$ values where the common modal projection operates adequately, delineating a region in the model parameter space ($k_2^p<1.0$ kN/mm) where the support points properly lead to interpolative CB matrices (Eq.~\ref{eq11}). Now, from Figure~\ref{F5}(b) it is clear that the threshold value of $k_2^p=1.0$ kN/mm identified in Figure~\ref{F5}(a) is related exclusively to the crossing of the $45^{\text{th}}$ and $46^{\text{th}}$ modes. In other words, the mode crossing does not depend on the reference configuration $\bfth^o$, but the ill-conditioning region does. This observation motivates the recomputation of Figure~\ref{F5}(a), but this time using a reference configuration $\bfth^o$ being larger than the $k_2$ threshold. Figure~\ref{F6} shows this case for $\bfth^o=[m,k_1,k_2]$, where $k_2=1.1$ kN/mm. These results show that the full rank and the ill-conditioned region flipped when compared against Figure~\ref{F5}(a). Based on these results, it is possible to argue that the ill-conditioning of the common modal projection can be controlled by selecting the reference configuration $\bfth^o$. 

\newpage
From these cases, it is possible to draw some insights about the role of the mode crossing over the well-conditioning of the common modal projection required in Eq.~(\ref{eq9}). The first observation to highlight is that swapping modes within the retained set does not introduce ill-conditioning in $\hat{\bfPhi}^{p,o}$. The problem of ill-conditioning arises when at least one of the swapping modes belongs to the retained set. The second observation is that crossing modes can trigger the ill-condition of $\hat{\bfPhi}^{p,o}$, but this ill-condition can be controlled by selecting a proper reference $\bfth^o$. Then, it is possible to combine multiple interpolation schemes, each one with its own reference, with the intention of extending the perturbation size. This situation is discussed later in Section 5. In any case, these observations introduce a new consideration in deciding the number of retained modes $q$ for modal reduction, traditionally chosen based exclusively on the accuracy of the reduced system to capture the whole system dynamics \cite{kim2015enhanced,krattiger2019interface}.

\section{Limits of the Model Parameter Space to Avoid Ill-conditioned Modal Projections}
\label{S:3}
This section presents a procedure to identify the model parameter space $\bfTh^{o}$ in which the model parameter $\bfth^p$ leads to well-conditioned $(\bfR^o)^T\bfPhi^p$ for a given reference configuration $\bfth^o$, i.e., $\{\bfth^p\in\bfTh^o|\det[(\bfR^o)^T\bfPhi^p]\neq0\}$. The notation of the model parameter space uses the superscript $o$ to indicate explicitly the dependency on the reference configuration $\bfth^o$, as it was shown in the previous section. The proposed procedure to identify $\bfTh^o$ is based on a multistage sampling approach described next.

\subsection{Sampling Scheme to Detect Ill-conditioned Modal Projections}
\label{S:3.1}

The general idea of the procedure is to define $\bfth^p$ through samples generated in regions around the nominal configuration $\bfth^o$. For each sample, the conditioning of $(\bfR^o)^T\bfPhi^p$ is checked, and the sample is labeled as "accepted" or "rejected". The sampling region is iteratively increased to cover the space of interest, allowing the identification of $\bfPhi^o$ as the region delimited by the samples labeled as "accepted". The detail of this general idea is presented below.

\begin{figure}[ht]
    \centering
    \includegraphics{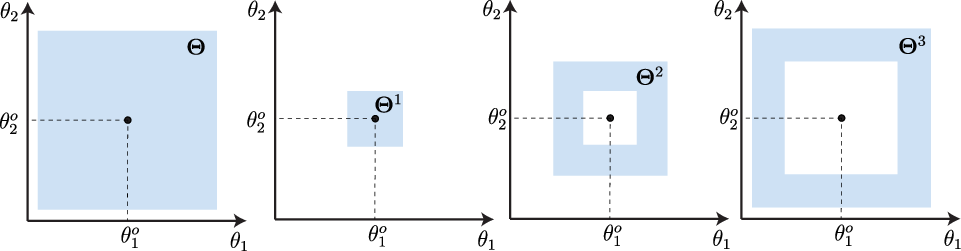}
    \caption{Model parameter space of interest $\bfTh$ and its mutually exclusive subdivisions (highlighted in blue). Example using two-dimensional model parameters and three subdivisions.}
    \label{F7}
\end{figure}

Let the model parameter vector $\bfth$ be defined in a two-dimensional space $\bfTh$, such that $\bfth=[\theta_1,\theta_2]\in\bfTh$. This selection is presented to facilitate the visualization of the implementation and does not represent a dimensionality constraint in the procedure. The model parameters of the reference substructure are denoted as $\bfth^o$, with components $\theta_1^o$ and $\theta_2^o$. First, the space of interest $\bfTh$ is divided into $n$ mutually exclusive subspaces $\{ \bfTh^i;i=1,...,n \}$, as it is represented in Figure~\ref{F7} for the particular case of a two-dimensional model parameter space and $n=3$. Figure~\ref{F7} first presents the interest model parameter space $\bfTh$ along with the reference model parameter $\bfth^o$, and subsequently, the subspaces $\{ \bfTh^i;i=1,...,n \}$. Here, different sampling strategies could be applied to populate these spaces \cite{helton2006survey,kleijnen2005overview}, with Latin Hypercube being one of the most widely adopted algorithms \cite{helton2003latin}. One strategy for sampling is the use of a Latin Hypercube to generate samples that fill the space of interest $\bfTh$. These samples are then relocated to their respective subspace $\bfTh^i$, which potentially leads to each subspace $\bfTh^i$ having a different number of samples $N_i$.

\begin{figure}[ht]
    \centering
    \includegraphics{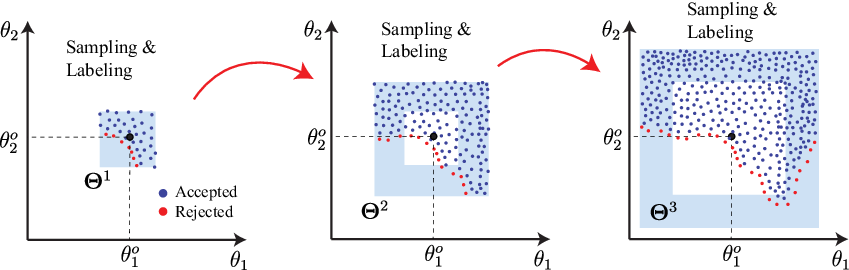}
    \caption{Schematic representation of the sampling-based identification of the suitable space for CB matrix interpolation, highlighting in blue the "accepted" and in red the "rejected" samples.}
    \label{F8}
\end{figure}

The samples are evaluated in sequence, starting from $\bfTh^1$ and progressing forward until $\bfTh^n$ as it is schematized in Figure~\ref{F8}. The conditioning of $(\bfR^o)^T\bfPhi^k$ is assessed for each sample, identified here as $k$, by computing its rank, determinant, or the conditioning number. The sample $k$ is labeled as "accepted" if $(\bfR^o)^T\bfPhi^k$ is well-conditioned, otherwise it is labeled as "rejected". As the first samples are close to the reference configuration $\bfth^o$, it is expected that the first samples will be labeled as "accepted". This evaluation process continues until either a sample is labeled as "rejected" or all samples have been processed. After rejecting the first sample, the subsequent process is modified. Now, before checking the conditioning of $(\bfR^o)^T\bfPhi^k$, the location of the sample $\bfth^k$ within the model parameter space is considered. If $\bfth^k$ is closer to a rejected sample than to an accepted one, its label is skipped and the following sample is evaluated. This is the reason why in Figure~\ref{F8} there are regions with no samples, i.e., unlabeled samples are not presented. This step reduces the need to compute $(\bfR^o)^T\bfPhi^k$ for samples that are already in the rejection region. Note that computing $\bfPhi^k$ is the most computationally expensive step since it involves the analysis of the $k$-th substructure using FEM. Also, this is the reason to divide the model parameter space in subspaces and evaluate samples from $\bfTh^1$ to $\bfTh^n$, since it facilitates the delimitation of the well-conditioning modal projections. There is a consideration that should be highlighted regarding the computation of the distance between samples. As the component of the model parameter may contain dissimilar orders of magnitude (e.g., length and thickness), it is a common practice to compute these distances in a normalized space \cite{fischer2023enhanced}. The normalization can be introduced based on maximum and minimum limits of the interest space $\bfTh$ or by transforming the samples into a standard normal space \cite{de2023choice}. Then, the distance between samples can be computed based on the $L_2$-norm. The algorithm concludes whether all subspaces have been explored or if a subspace $\bfTh^i$ yields zero accepted samples. The main steps of this algorithm are presented in Algorithm~\ref{alg:1}.

\begin{algorithm}
\caption{Sampling Generation and Labeling}\label{alg:1}
\begin{algorithmic}[1]
\State Define $\bfth^o$ and compute $\bfR^o$
\State Divide $\bfTh$ in $n$ subspaces
\State Set the number of samples per subspace $\{N_i;i=1,..., n\}$
\State $k \gets 0$  \Comment{Number of labeled samples}
\For{$i=1:n$}
    \For{$j=1:N_i$} 
        \State Sample $\tilde{\bfth}\in\bfTh^i$
        \If{No samples have been rejected}
            \State Perform Algorithm~\ref{alg:2}
        \Else
            \State $\bfth^* \gets$ closest sample between $\tilde{\bfth}$ and $\{\bfth^p;p=1,...,k\}$
            \If{$\bfth^*$ has label "Accepted"}
                \State Perform Algorithm~\ref{alg:2}
            \EndIf
        \EndIf
    \EndFor
    \State Terminate algorithm if no samples are accepted in $\bfTh^i$
\EndFor
\end{algorithmic}
\end{algorithm}

\begin{algorithm}
\caption{Checking Ill-conditioned Modal Projections}\label{alg:2}
\begin{algorithmic}[1]
\State Compute $\tilde{\bfPhi}$ for $\tilde{\bfth}$
\If{$\bfR^o\tilde{\bfPhi}$ is well-conditioned}
    \State Label sample as "Accepted"
\Else
    \State Label sample as "Rejected"
\EndIf
\State $k \gets k+1$
\State $\bfth^k \gets \tilde{\bfth}$
\end{algorithmic}
\end{algorithm}

\newpage
\subsection{Implementation of Interpolative Schemes}
\label{S:3.2}

\begin{figure}[ht]
    \centering
    \includegraphics{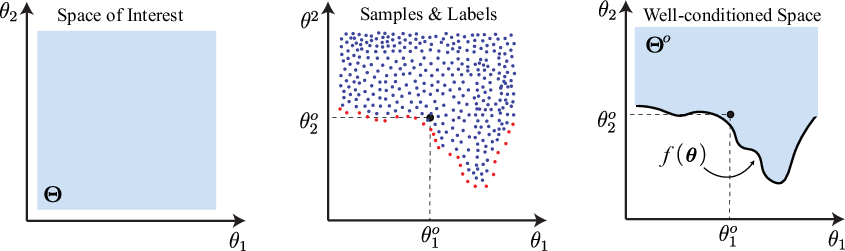}
    \caption{Identification of the well-conditioning modal projection boundary based on samples and support vector machines.}
    \label{F9}
\end{figure}

Considering the previous sampling process and its binary labeling, it is possible to employ a binary classification model based on Support Vector Machines (SVM) for defining the nonlinear separation boundary between two groups \cite{kecman2005support}. This process is represented in Figure~\ref{F9}, where the first plot represents the space of interest $\bfTh$, the second plot presents the labeled samples obtained applying the procedure described in Section 3.1, and the third plot represents the model parameter space $\bfTh^o$ that is suitable to be used in the support point definitions for any desired interpolation scheme. The SVM allows two main things: (1) the detection of the boundary described by $f(\bfth)$, and (2) the labeling of any new model parameter $\bfth$ to identify if it is inside or outside $\bfTh^o$.

\begin{figure}[ht]
    \centering
    \includegraphics{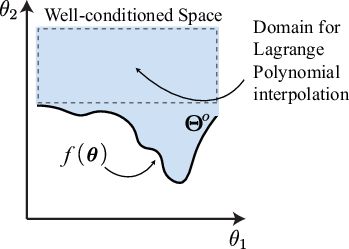}
    \caption{Maximum usable region for Lagrange Polynomial interpolation schemes.}
    \label{F10}
\end{figure}

In the work presented by Mencik \cite{mencik2024improved}, the interpolation is formulated using Lagrange Polynomials, as described in the second block of Figure~\ref{F2}. Under this scheme, the distribution of the support points $\bfth^p$ (which are required for computing the reduced CB matrices in Eq.~\ref{eq11}) inherently constrains the usable interpolative domain to a hypercube. In scenarios such as the one depicted in Figure~\ref{F9} (where the boundary of $\bfTh^o$ is not parallel to the main model parameter axes), such implementations significantly restrict the effective interpolation region. For illustration, considering the $\bfTh^o$ region from Figure~\ref{F9}, the maximum domain where a Lagrange Polynomial interpolation can be effectively implemented is shown by dashed lines in Figure~\ref{F10}; it is evident that this region is substantially smaller than $\bfTh^o$. Later in Section 6, an illustrative example is presented to highlight a situation like this explicitly.

Strictly speaking, the entirety of the domain $\bfTh^o$ can be utilized for implementing an interpolation technique. To circumvent the limitations associated with Lagrange Polynomials, an alternative approach is proposed, leveraging the samples already available from Section 3.1 in conjunction with the SVM model. In this context, the proposal corresponds to employ the input data $\{\bfth^p;\,p=1,...,k\}$ and their corresponding outputs $\{ \hat{\bfM}^{p,o},\hat{\bfK}^{p,o},\hat{\bfF}^{p,o};\,p=1,...,k\}$ to train an interpolation model based on Kriging \cite{lophaven2002dace}. This adoption offers two primary benefits: (1) the necessary support points are already available from the previous analysis, and (2) the SVM provides a robust discriminant to ascertain whether any new model parameter $\bfth$ lies within $\bfTh^o$. Consequently, this approach enables the establishment of interpolations to accurately estimate $\hat{\bfM}(\bfth)$, $\hat{\bfK}(\bfth)$, and $\hat{\bfF}(\bfth)$ for any $\bfth\in\bfTh^o$, thereby allowing the utilization of the whole domain $\bfTh^o$.

\subsection{Description of the Proposed Interpolation Approach}
\label{S:3.3}

Although Kriging interpolation has the potential to predict the CB matrices $\hat{\bfM}(\bfth)$, $\hat{\bfK}(\bfth)$, and $\hat{\bfF}(\bfth)$, a significant computational cost may arise due to the high dimensionality of the output space \cite{ji2023efficient}. Therefore, in this section a dimensionality reduction is proposed over $\hat{\bfM}^{p,o},\hat{\bfK}^{p,o},$ and $\hat{\bfF}^{p,o}$ in the form of Principal Component Analysis (PCA) before training the Kriging interpolation model. This technique was chosen for its ability to reduce the original output space into a lower-dimensional latent space by leveraging linear correlation between the output space features \cite{abdi2010principal}. Noting that PCA is a linear transformation, this latent space retains the interpolative capabilities with respect to the model parameters.

Let each entry of $\hat{\bfM}^{p,o},\hat{\bfK}^{p,o},$ and $\hat{\bfF}^{p,o}$ (Eq.~\ref{eq11}) constitute a feature in the PCA, such that a vector of features $\bfX^{p,o}$ for a particular set of model parameters $\bfth^p$ can be represented as in Eq.~(\ref{eq12}):
\begin{equation}
    \bfX^{p,o} = [\hat{\bfM}^{p,o}_1,...,\hat{\bfM}^{p,o}_r,\hat{\bfK}^{p,o}_1,...,\hat{\bfK}^{p,o}_r, (\hat{\bfF}^{p,o})^T ],
    \label{eq12}
\end{equation}

\noindent where $\hat{\bfM}^{p,o}_m$ and $\hat{\bfK}^{p,o}_m$ corresponds to the $m$-th row of $\hat{\bfM}^{p,o}$ and  $\hat{\bfK}^{p,o}$, respectively. Here, $r$ denotes the total number of DoF in the reduced system, i.e., the size of $\hat{\bfM}^{p,o}$ and $\hat{\bfK}^{p,o}$. Subsequently, the matrix of data points $\bfX^{data}$ for the PCA training can be constructed by appending each vector of features $\bfX^{p,o}$ for all support points $p=1,...,k$ as shown in Eq.~(\ref{eq13}):
\begin{equation}
    \bfX^{data} = 
    \begin{bmatrix}
        \bfX^{1,o} \\
        \vdots \\
        \bfX^{k,o}
    \end{bmatrix}.
    \label{eq13}
\end{equation}

The stack of features and support points results in a matrix $\bfX^{data}$ of size $k\times[2r+1]r$. Subsequently, and following the standard implementation of PCA \cite{abdi2010principal}, a transformation matrix $\bfQ^o$ of size $[2r+1]r\times u$ can be obtained to project $\bfX^{p,o}$ into a latent space of dimension $u$, resulting in a vector of latent features $\bfY^{p,o}$ of size $k\times u$ with $u\ll r$ computed following Eq.~(\ref{eq14}):
\begin{equation}
    \bfY^{p,o} = [\bfX^{p,o}-\bar{\bfX}^{data}]\bfQ^o.
    \label{eq14}
\end{equation}

tiHere, $\bar{\bfX}^{data}$ refers to a vector of mean values for each feature in $\bfX^{data}$. The selection of the number of latent features to retain $u$ depends primarily on the error introduced in the prediction of the FRF, while the identification of $\bfQ^o$ is established by the Singular Value Decomposition of $\bfX^{p,o}-\bar{\bfX}^{data}$ \cite{abdi2010principal}. After the PCA reduction is implemented, a Kriging interpolation scheme can be deployed between the input set $\{\bfth^p;\,p=1,...,k\}$ and the new set of outputs $\{ \bfY^{p,o};\,p=1,...,k\}$. Then, for any new sample $\bfth\in\bfTh^o$, its corresponding vector of latent features $\bfY^{o}$ can be estimated using the Kriging interpolation, i.e.,$\bfY^{o}(\bfth)$. These latent features $\bfY^{o}$ can be transformed into the initial space by computing Eq.~(\ref{eq15}):
\begin{equation}
    \bfX^{o}(\bfth) \approx \bfY^{o}(\bfth)[\bfQ^o]^T  + \bar{\bfX}^o
    \label{eq15}.
\end{equation}

\begin{figure}[ht]
    \centering
    \includegraphics{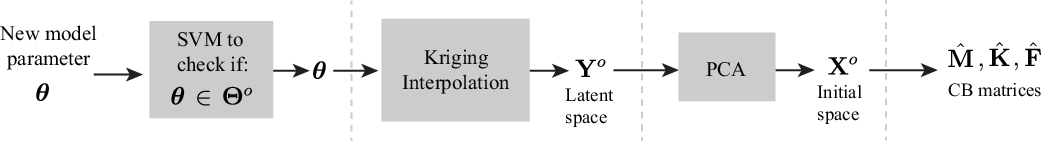}
    \caption{Schematic representation of the proposed PCA/Kriging interpolation.}
    \label{F11}
\end{figure}

Then, its CB reduced matrices can be recovered from $\bfX^{o}$ following the same arrangement of features presented in Eq.~(\ref{eq12}). A schematic of the proposed interpolation workflow is presented in Figure~\ref{F11}. Please note that although the Kriging/PCA scheme is new for metamaterial applications, similar approaches have been previously implemented for natural hazard assessment \cite{jia2013kriging}, reacting flow analysis \cite{aversano2019application}, and biomechanical applications \cite{steer2020predictive}. 

Ultimately, when comparing the resulting interpolation strategy (pairing SVM, PCA, and Kriging) against a  Lagrange Polynomials interpolation, several pros and cons arise. First, the proposed interpolation is capable of operating within the whole well-conditioned space $\bfTh^o$, while Lagrange Polynomials are restricted to subregions of $\bfTh^o$ as depicted in Figure~\ref{F11}. Second, the computational cost to generate support points in the proposed scheme is more demanding compared to Lagrange Polynomials. However, this will depend on the dimensionality of the model parameter space and the polynomial degree chosen. For large dimensionalities, the Lagrange Polynomials could also become computationally expensive since they grow as $p^n$ (assuming equal polynomial degree for all dimensions), where $n$ is the number of dimensions in the model parameter space and $p$ is the number of support points per dimension (e.g., a second-order interpolation requires $p=3$ support points). Third, the proposed interpolation can be modified to extend the interpolative region beyond $\bfTh^o$, while the Lagrange Polynomial only admits this when the boundaries of $\bfTh^o$ are parallel to the model parameter axes. This particular capability is fully described next in Section~\ref{S:4}, and then in the second illustrative example in Section~\ref{S:6}.

\section{Extending the Interpolative Space}
\label{S:4}

The previous procedure is now modified to extend the interpolative space by selecting different reference sets of model parameters. The modification is supported by the evidence provided previously in Figures~\ref{F5} and \ref{F6}, where different selections of $\bfth^o$ lead to different well-conditioned regions. Then, the general idea is to identify these regions and their reference set of model parameters to train an interpolative model that can be used within each region. Afterward, the respective interpolative model will be chosen depending on the region where the new model parameter is located. 

\begin{figure}[ht]
    \centering
    \includegraphics{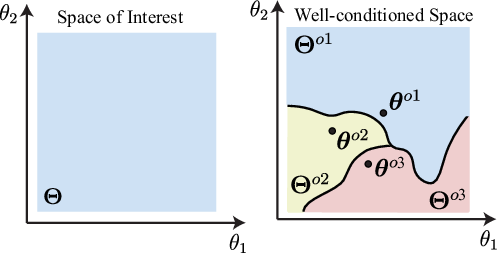}
    \caption{Schematic representation of subdomains where the common modal projections are well-conditioned. Each subdomain has its own reference set of model parameters.}
    \label{F12}
\end{figure}

Before presenting the procedure, a brief description of the nomenclature is needed. Let Figure~\ref{F9} represent the model parameter space $\bfTh$ of the substructure that we want to employ. Based on the procedure described in Section~\ref{S:3}, it is possible to identify the region $\bfTh^o$ where the modal projection is well-conditioned. For that purpose, a reference set of model parameter $\bfth^o$ is needed to compute $\bfR^o$ as described in Eq.~(\ref{eq8}), such that the modal projection for the $p$-th support point (denoted as $\bfth^p$) is well-conditioned, i.e., the product $(\bfR^o)^T\bfPhi^p$ shown in Eq.~(\ref{eq9}) must be invertible to compute $\hat{\bfPhi}^{p,o}$. Something important to note is that the identification of $\bfTh^o$ is independent of the substructure $\bfth^o$ used as reference, as long as $\bfth^o\in \bfTh^o$. Now, consider the same model parameter space $\bfTh$ being divided by $m$ non-overlapping regions $\{\bfTh^{o1},\bfTh^{o2}...,\bfTh^{om}\}$, each one containing its own reference model parameter $\{\bfth^{o1},\bfth^{o2}...,\bfth^{om}\}$ that leads to own modal projections $\{\bfR^{o1},\bfR^{o2}...,\bfR^{om}\}$. The modal projection for the $p$-th support point is denoted as $\hat{\bfPhi}^{p,o1}$ if it belongs to the first region $\bfTh^{o1}$, as $\hat{\bfPhi}^{p,o2}$ if it belongs to the second region $\bfTh^{o2}$, and so on. In this way, there are two counters: the number of regions (and reference substructures) and the number of support points. This division of the model parameter space is exemplified in Figure~\ref{F12} by using three subdomains.

Something important to remark is that the identification of $\bfTh^{om}$ is independent of the substructure $\bfth^{om}$ used as reference, as long as $\bfth^{om}\in \bfTh^{om}$. Thus, any model parameter configuration $\bfth$ can be used as the reference substructure as long as it belongs to the region whose boundary is being tried to identify. Let's consider the two-dimensional model parameter space presented in Figure~\ref{F12} as the base case to explain the procedure to identify the well-conditioned regions. The algorithm starts by generating $N$ samples in $\bfTh$ favoring space-filling algorithms, such as Latin Hypercube Sampling \cite{florian1992efficient}. This set of samples corresponds to $\{\bfth^p;\,p=1,...,N\}$. Then, the FE matrices corresponding to each of the generated samples are computed $\{ {\bfM}^{p},{\bfK}^{p},{\bfF}^{p};\,p=1,...,N\}$, followed by their respective fixed-interface modes $\{ \bfPhi^p;\,p=1,...,N \}$. The first sample $\bfth^1$ is set as the reference configuration $\bfth^{o1}$ for the first region $\bfTh^{o1}$, and its modal projection is defined as $\bfR^{o1}=\bfM_{jj}^{o1}\bfPhi^{o1}$. The subsequent samples $\bfth^p$ are incorporated to check if the product $(\bfR^{o1})^T\bfPhi^p$ is well-conditioned. All samples that are well-conditioned will be tagged to belong to $\bfTh^{o1}$ until reaching the first sample $p=l$ that yields an ill-conditioned projection. Now, this sample $\bfth^l$ will be taken as the reference set $\bfth^{o2}$ for the second region $\bfTh^{o2}$, such that the new common projection corresponds to $\bfR^{o2}=\bfM_{jj}^{o2}\bfPhi^{o2}$.

\begin{figure}[ht]
    \centering
    \includegraphics{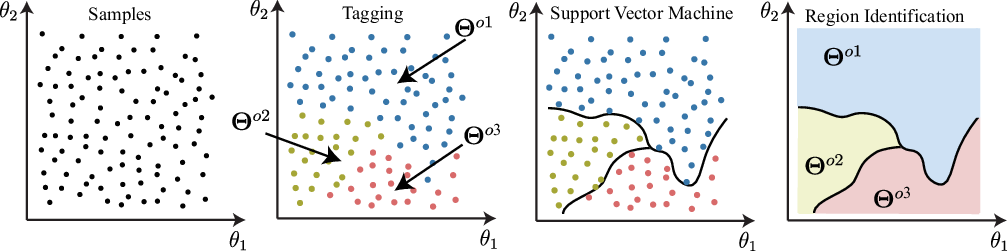}
    \caption{Scheme for the identification of well-conditioned modal projection regions. Three main stages are presented: sampling generation, tagging, and classification via Support Vector Machines.}
    \label{F13}
\end{figure}

Then, any subsequent sample must check if $(\bfR^{o1})^T\bfPhi^p$ or $(\bfR^{o2})^T\bfPhi^p$ is well-conditioned, tagging the sample to the corresponding region. If the sample leads to an ill-conditioned projection in both cases, another reference set is established (in this case $\bfth^{o3}$) together with a new region $\bfTh^{o3}$. This procedure continues until all samples are tagged, leading to tagged samples like the ones presented in Figure~\ref{F13}. The algorithm for tagging the samples is presented in Algorithm~\ref{alg:3}. After tagging all samples, it is possible to adopt an SVM to identify the boundaries of each region and to tag any new set of model parameters in its corresponding region. These types of problems are traditionally solved via decomposition strategies where the multiclass problem is divided into multiple binary classification problems as described by Lorena et al. \cite{lorena2008review}.

\begin{algorithm}
\caption{Tagging Sequence}\label{alg:3}
\begin{algorithmic}[1]
\State Generate samples: $\{ \bfth^p;\,p=1,...,N \}\in\bfTh$ \Comment{$N$ samples are generated}
\For{$k=1:N$}
    \State Compute $\bfM^p$, $\bfK^p$, $\bfF^p$
    \State Compute $\bfPhi^p$
\EndFor
\State $m\gets 1$ \Comment{Counter for number of regions}
\State $\bfth^{om}\gets\bfth^{1}$ \Comment{First sample is set as reference}
\State Define $\bfR^{om}=\bfM_{jj}^{om}\bfPhi^{om}$ based on $\bfth^{om}$
\State Tag $\bfth^1$ as $m$
\For{$p=2:N$}
    \For{$i=1:m$}
        \If{$(\bfR^{oi})^T\bfPhi^p$ is well-conditioned}
            \State Tag $\bfth^p$ as $m$
        \EndIf
    \EndFor
    \If{No label is assigned to $\bfth^p$}
        \State $m \gets m+1$
        \State $\bfth^{om}\gets\bfth^{p}$ 
        \State Define $\bfR^{om}=\bfM_{jj}^{om}\bfPhi^{om}$ based on $\bfth^{om}$
        \State Tag $\bfth^p$ as $m$
    \EndIf
\EndFor    
\end{algorithmic}
\end{algorithm}

\begin{figure}[ht]
    \centering
    \includegraphics[width=0.9\linewidth]{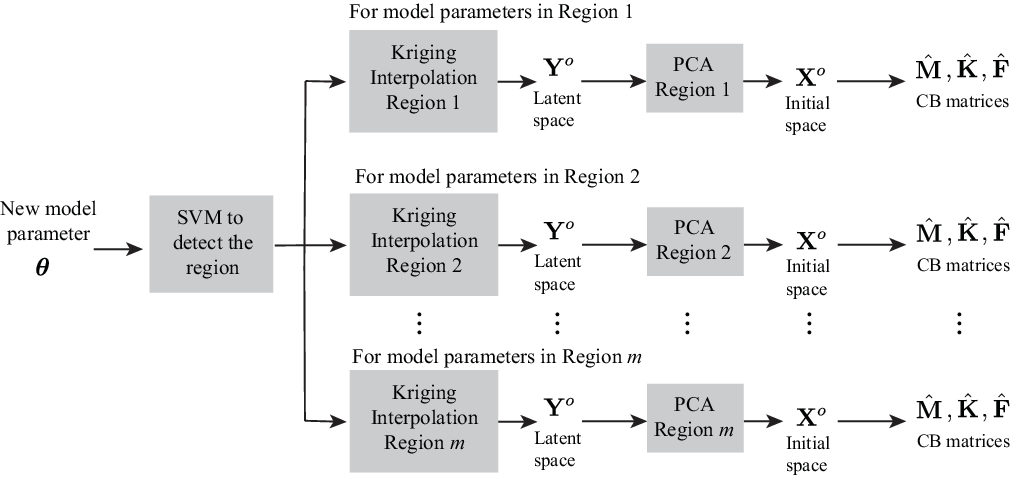}
    \caption{General schematic of the region-based surrogate model.}
    \label{F14}
\end{figure}

Now, $m$ interpolation models are trained following the same scheme presented in Section~\ref{S:3.3}, i.e., each region identified previously will be associated with a unique surrogate model. Then, any new sample $\bfth$ is evaluated first by the SVM to identify the region to which it belongs, and then the respective surrogate model is used to predict $\hat{\bfM}$, $\hat{\bfK}$, and $\hat{\bfF}$. The scheme of this procedure is presented in Figure~\ref{F14}.

\section{Illustrative Example: Unit Cell with Square Plate and Circular Core}
\label{S:5}
    
\subsection{Nominal Structure}
\label{S:5.1}

\begin{figure}[ht]
    \centering
    \includegraphics[width=1\linewidth]{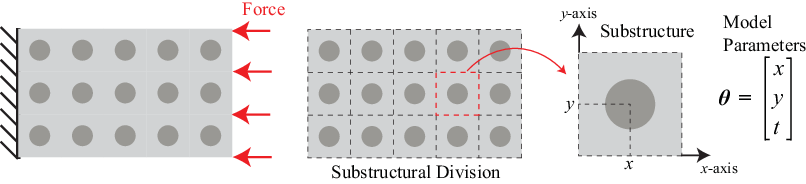}
    \caption{Benchmark problem used for validation taken from \cite{mencik2024improved}.}
    \label{F15}
\end{figure}

The proposed method is implemented over a benchmark problem taken from \cite{mencik2024improved} for validation purposes. The main structure is shown in Figure~\ref{F15}, comprising the assembly of $5\times3$ identical square plates (substructures), each of them with a thick circular core. The structure is fixed at its left boundary, and a unit force is applied horizontally over four points on the right boundary, as illustrated also in the figure. Both the plate and the circular core are modeled as linear elastic materials, with properties presented in Table~\ref{T1}. At the same time, a proportional damping is applied, adopting $\alpha=0.01\ \text{s}^{-1}$ (mass proportional coefficient) and $\beta=10^{-8}\ \text{s}$ (stiffness proportional coefficient). Note that this example corresponds to an in-plane vibration problem, where the structure's response is studied by computing the average of the quadratic velocity FRF associated with all interface degrees of freedom (dashed lines in Figure~\ref{F15}) in the $x$-direction. In this case, the model parameter vector corresponds to the geometrical features that describe the circular core, such that $\bfth=[x,y,t]$, being $x$ and $y$ the coordinates of the circular core center, and $t$ its thickness. The plate corresponds to a $200\times200$ mm square with a thickness of 1 mm, while the circular core has a diameter of 100 mm and a thickness of 5 mm, and it is located at the centroid of the square. 

\begin{table}
    \centering
\caption{Material properties for the matrix and the circular core.}
\label{T1}
    \begin{tabular}{lccc}
        \hline
        & $E$ [GPa] & $\nu$ & $\rho$ [kg/$\text{m}^3$]\\ \hline
        Matrix & 70 & 0.35 & 2700\\
        Circular Core & 340 & 0.27 & 19250\\
        \hline
    \end{tabular}
\end{table}

\begin{figure}[ht]
    \centering
    \includegraphics[width=0.8\linewidth]{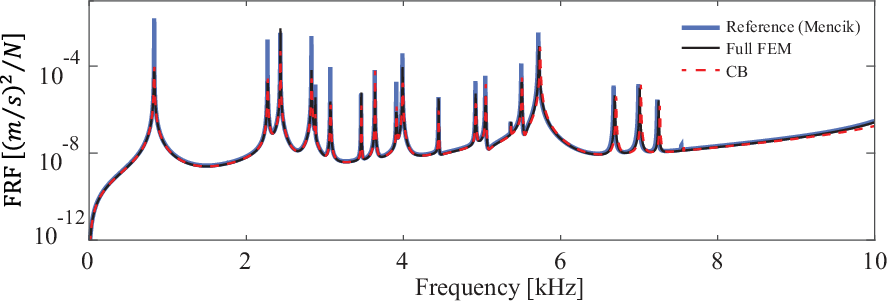}
    \caption{FRF of the structure with nominal model parameters, showcasing the benchmark results from \cite{mencik2024improved} in blue-solid lines, the full FE results in black-solid lines, and the CB-reduced results as red-dashed lines.}
    \label{F16}
\end{figure}

Figure~\ref{F16} presents the FRF obtained by adopting different models in the frequency range of 0-10000~Hz.  The first FRF corresponds to the structural response extracted directly from \cite{mencik2024improved} (the respective author gently provided the data of the plot). The second FRF corresponds to the response obtained by solving the problem in ANSYS and using PLANE182 linear triangular elements, which correspond to plate elements with two DoF per node (displacements in the x and y-directions). The third FRF corresponds to the implementation described in Section~\ref{S:2.2}, where the substructure mass and stiffness matrices were obtained using ANSYS MAPDL (using the same type of elements described before) and exported to MATLAB to obtain the reduced matrices (Eq.~\ref{eq11}) and compute the FRF. The baseline mesh comprises 717 nodes, including 64 boundary nodes. As all substructures are identical, the reduced matrices (Eq.~\ref{eq11}) are computed setting $\bfth=\bfth^p=\bfth^o$. For the modal projection, $q=3$ fixed-interface DoF were retained. Ultimately, the three methods align closely, with negligible deviations between the benchmark and full FE results attributable to mesh differences.

\subsection{Analysis of Aperiodic Substructures.}
\label{S:5.2}

\begin{figure}[ht]
    \centering
    \includegraphics[width=1\linewidth]{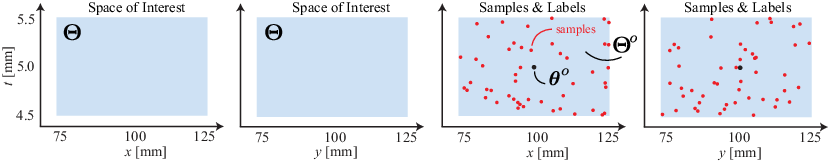}
    \caption{Samples used to detect the well-conditioned region $\bfTh^o$ and train the proposed surrogate model.}
    \label{F17}
\end{figure}

The interest of this section is to show the performance of the proposed method (Section~\ref{S:3}) when a structure contains non-identical substructures, i.e., each substructure is considered perturbed with respect to the nominal configuration. The space of interest for the model parameters $\bfth=[x,y,t]\in\bfTh$ corresponds to $x=[75,125]$ mm, $y=[75,125]$ mm, and $t=[4.5,5.5]$ mm. The reference set of model parameters $\bfth^o$ is taken as the nominal values for these parameters, i.e., $x=100$ mm, $y=100$ mm, and $t=5$ mm. As an initial test, 50 samples ($k=50$) were generated in $\bfTh$ using LHC sampling and subsequently labeled using Algorithm~\ref {alg:1}. In this process, ANSYS MAPDL was used exclusively to obtain the mass and stiffness matrices of each sample, and then exported to MATLAB to compute all required steps in Algorithms \ref{alg:1} and \ref{alg:2}. After completing the iterative labeling process, it was observed that all modal projections were well-conditioned, indicating the existence of a single region within the model parameter space (i.e., all model parameters within the space have a well-conditioned projection). These samples are presented in Figure~\ref{F17} and then used to train a PCA model to reduce the dimension of $\bfX^{p,o}$  (size around 30,000) down to 6 for the latent features $\bfY^{p,o}$ following the method described in Section~\ref{S:3.3}. Then, the Kriging model was trained using the same 50 samples to create a mapping between the model parameter $\bfth$ and the latent feature space $\bfY^o$, achieving median relative errors of less than 5\% (obtained through a leave-one-out cross-validation \cite{pang2023enhanced}). At this point, the scheme presented in Figure~\ref{F11} is ready for use. 

\begin{figure}[ht]
    \centering
    \includegraphics[width=1\linewidth]{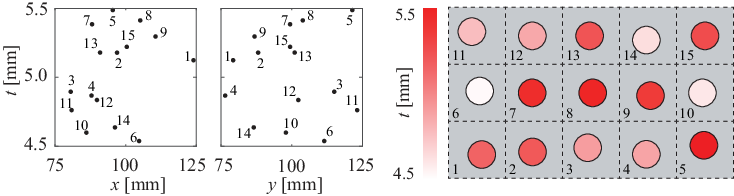}
    \caption{Model parameters selected for computing the FRF within the design space and schematic representation of the structure evaluated.}
    \label{F18}
\end{figure}

\begin{figure}[ht]
    \centering
    \includegraphics[width=0.8\linewidth]{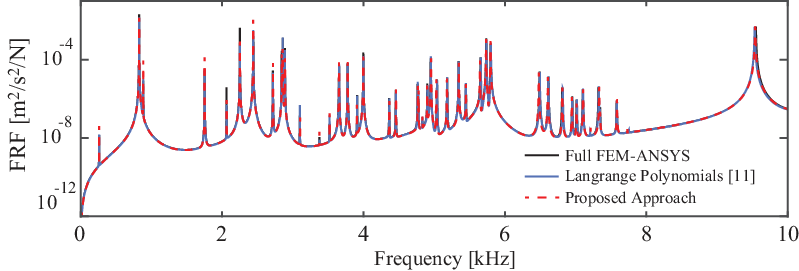}
    \caption{FRF of the structure with random model parameters, showcasing the results using a Lagrange polynomial-based interpolation in blue-solid lines, the exact CB results in black-solid lines, and the proposed PCA-Kriging surrogate prediction in red-dashed lines.}
    \label{F19}
\end{figure}

The efficacy of the proposed procedure is assessed by assembling 15 substructures, whose model parameters were randomly sampled and previously established in \cite{mencik2024improved}. The model parameters of the selected substructures are shown in Figure~\ref{F18}. The FRF of the assembly is computed using three methods. The first method corresponds to a high-fidelity FEM based exclusively on ANSYS. The second method corresponds to a three-dimensional Lagrange polynomial interpolation scheme with 2nd order polynomials, which follows the procedures in \cite{mencik2024improved}. Gaussian points are defined at $\eta_1=h[1-\sqrt{5/3}P]$, $\eta_2=h$, and $\eta_3=h[1+\sqrt{5/3}P]$, where $h$ corresponds to the nominal value for each perturbed parameter (i.e., $x=100$ mm, $y=100$ mm, $t=5$ mm,) and $P$ refers to the perturbation magnitude (i.e., 0.25). This implementation requires only $3\times3\times3=27$  support points. Finally, the third method corresponds to the proposed method presented in Section~\ref{S:3.3}. The resulting FRFs are presented in Figure~\ref{F19}, verifying the ability of the proposed method to predict the dynamic response of aperiodic structures. Despite the good match between the methods, the proposed method has the drawback of requiring more support points, 50, compared to the 27 points used by the Lagrange interpolation. However, the drawback of the Lagrange interpolation lies in the fact that it does not check if the new sample will have a well-conditioned projection. As this example contains only a single region where the model parameters are well-conditioned, the benefit of the proposed method is not as significant as the example presented next.

\section{Illustrative Example: Beam-like Structure with Vibration Attenuation Bands.}
\label{S:6}

The motivation of this section is to show a comparative study between the proposed interpolation strategy highlighted in Section~\ref{S:4} and the Lagrange interpolation approach presented in \cite{mencik2024improved} when the geometrical variations lead to multiple well-conditioned projection regions. 
    
\subsection{Nominal Structure}
\label{S:6.1}

\begin{figure}[ht]
    \centering
    \includegraphics[width=0.5\linewidth]{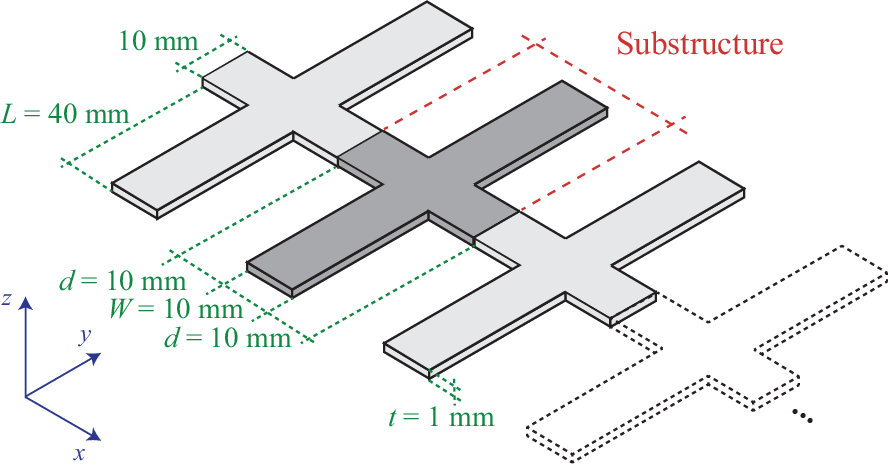}
    \caption{Scheme of the metamaterial assembly composed of 10 substructures supported in fixed-free condition.}
    \label{F20}
\end{figure}

The example corresponds to a 2D beam-like structure composed of 10 substructures as shown in Figure~\ref{F20}. Substructures comprise two cantilever plates that act as resonators, inducing band gaps at the structure level. The motivation to present this example is that this structure was previously studied in \cite{pereira2025vibration} to highlight the effect of geometrical variations on the frequency attenuation band so that it can serve as a benchmark problem in the field of dynamic and mechanical metamaterials. The structure vibrates out-of-plane ($z$-direction), and the material is considered linear elastic with elastic modulus $E=205$ GPa, density $\rho=7890\ \text{kg/m}^3$, and Poisson's ratio $\nu=0.3$. The geometrical characteristic of the nominal configuration is also presented in Figure~\ref{F20}, where the model parameter vector corresponds to $\bfth=[L,W]$. The FEM model is performed in ANSYS MAPDL using SHELL181 quadrilateral elements, consisting of four nodes with 3 DoF each (displacements in the $z$-direction and rotations in the $x$ and $y$ directions), while proportional damping is imposed at the substructural level to keep damping ratios of 0.01\% for the first two free-free vibration modes. The baseline mesh consisted of 329 nodes with 14 boundary nodes. The structure is fixed at its left end, and the response is studied by computing the FRF, defined here as the absolute deflection of the right end divided by a vertical displacement imposed at the left end. More details of the FEM modeling can be found in \cite{pereira2025vibration}.

\begin{figure}[ht]
    \centering
    \includegraphics[width=0.8\linewidth]{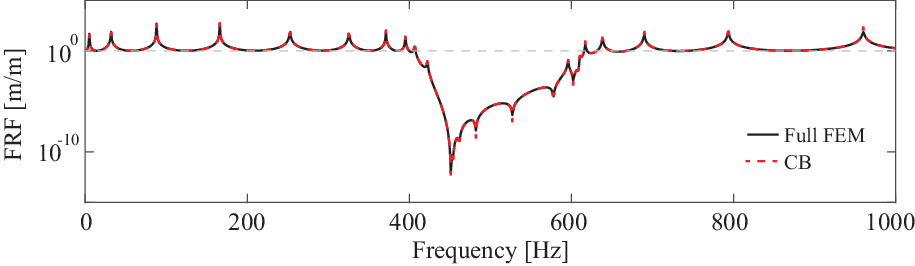}
    \caption{FRF of the structure with nominal model parameters, showcasing the FEM results in black-solid lines and the CB-reduced results as red-dashed lines.}
    \label{F21}
\end{figure}

The response of the nominal structure is obtained and presented in Figure~\ref{F21}, where two models are adopted: (1) a FEM implemented in ANSYS after a carefully verification via converge analysis, and (2) the CB model described in Section~\ref{S:2.2} adopting $\bfth=\bfth^p=\bfth^o$, similar to the example presented in the previous section and retaining ten fixed-interface vibration modes ($q=10$). As it is observed, the FEM and CB model match, indicating that the number of fixed-interface retained modes is enough to determine the vibration response of the structure.

\subsection{Training the Proposed Interpolation Method}
\label{S:6.2}

\begin{figure}[ht]
    \centering
    \includegraphics[width=0.9\linewidth]{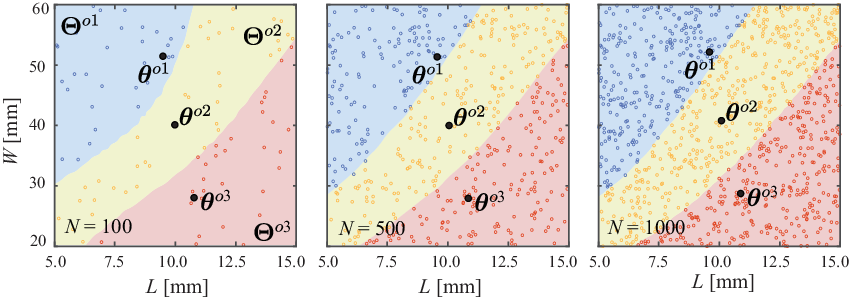}
    \caption{Implementation of Algorithm~\ref{alg:3} to detect well-conditioned projection regions using different numbers of samples: $N=100$, $N=500$, and $N=1000$. Three regions are detected, samples are shown in hollow circles, and reference model parameters used in each region are shown in black dots.}
    \label{F22}
\end{figure}

The parameter space $\bfTh$ is defined to cover variations of 50\% over the nominal length $L$ and width $W$ described in Figure~\ref{F20}. More specifically, the modal parameter space corresponds to $L=[5,15]$ mm and $W=[20,60]$ mm. The first step in the proposed method corresponds to the implementation of Algorithm~\ref{alg:3} (described in Section~\ref{S:4}) to identify the total number of well-conditioned projection regions. In this case, a total of 3 regions is observed, corresponding to $\bfTh^{o1}$, $\bfTh^{o2}$, and $\bfTh^{o3}$. The regions detected are presented in Figure~\ref{F22} after running Algorithm~\ref{alg:3} with different number of samples $N=100$, $N=500$, and $N=1000$. The samples are generated using LHC and presented as hollow circles over each of the three detected regions (highlighted with different colors). The reference model parameter for each region is also included. As expected, the quality of the SVM improves as more samples are used. 

Subsequently, PCA models were trained within each region following the directions presented in Section~\ref{S:3.3}. Here, the input feature described in Eq.~(\ref{eq12}) corresponds to each entry of the matrices $\hat{\bfM}^{p,o},\hat{\bfC}^{p,o},\hat{\bfK}^{p,o}$ and $\hat{\bfF}^{p,o}$. The precision of the PCA is evaluated by computing the reconstruction error, which is defined as the maximum relative error among the highest five free-free natural frequencies of the substructure following Eq.(\ref{eq16}). Here, $f_i$ corresponds to the frequencies obtained using the matrices $\hat{\bfM}^{p,o},\hat{\bfK}^{p,o}$ associated to the samples, while  $\hat{f_o}$ refers to the frequencies computed using the recovered matrices from Eq.(\ref{eq15}). The training was repeated for the three sets studied ($N=100, 500,$ and $100$ samples) to explore the effect of the sampling size on the quality of the PCA. For the three sample sets, reconstruction errors were lower than 0.1\% when a total of 10 latent features were retained, indicating that the PCA model does not require a large number of samples for its training. Then, the matrix $\bfX^{p,o}$  (size 8216) is reduced to 10 latent features ($\bfY^{p,o}$) following the method described in Section~\ref{S:3.3}.
\begin{equation}
    e_i^{PCA} = \left|\frac{f_i-\hat{f_i}}{f_i}\right|\times100.
    \label{eq16}
\end{equation}

Finally, three region-based Kriging interpolation models were trained using the aforementioned sample sets. The accuracy of the Kriging scheme is evaluated following a leave-one-out cross-validation approach. For the 100-sample set, median errors above 50\% were registered, dropping down to 2.2\% and 0.5\% for the 500 and 1000-sampling sets, respectively. As expected, the accuracy of the Kriging model improves as the number of support points increases. In contrast to the PCA, the Kriging models do benefit from a larger sample size during training. Given the lower error exhibited by the Kriging trained with $N=1000$, it is decided to use it in the subsequent analysis.

\subsection{Study of Aperiodic Structures}
\label{S:6.3}

\begin{figure}[ht]
    \centering
    \includegraphics[width=0.9\linewidth]{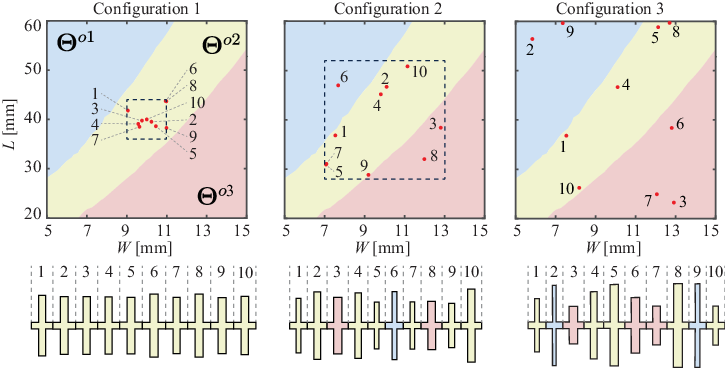}
    \caption{Aperiodic configurations studied. Samples used in each configuration are labeled and identified in the model parameter space with red circles. Bottom figures show a top view ($x-y$ plane) of each configuration. A specific color is used to identify each substructure with its respective region.}
    \label{F23}
\end{figure}

The capacity of the proposed interpolation model to identify the FRF is tested in three aperiodic structures. Each aperiodic structure follows the model depicted in Figure~\ref{F20}, but this time, a perturbation is randomly (and independently) introduced in each substructure. Each aperiodic structure presents different perturbation levels: 10\%, 30\%, and 50\% for the geometrical parameters $W$ and $L$. The scheme of these aperiodic structures is presented in Fig~\ref{F23}. The figure shows: (1) the pairs of $W$ and $L$ for each substructure, (2) the region indicating the perturbation level (dashed lines), and (3) the well-conditioned projection regions identified previously. From these three aperiodic configurations, it is clearly observed that Configuration 1 contains all $W-L$ pairs inside $\bfTh^{o2}$, while Configurations 2 and 3 also contain $W-L$ pairs in $\bfTh^{o1}$ and $\bfTh^{o3}$.

\begin{figure}[ht]
    \centering
    \includegraphics[width=0.3\linewidth]{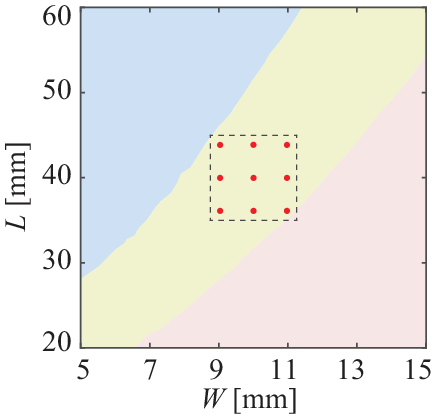}
    \caption{Selected support points for the two-dimensional Lagrange Polynomial interpolation scheme. Dashed lines indicate the design space region supported by the interpolation.}
    \label{F24}
\end{figure}

Alongside the proposed model, a Lagrange Polynomials interpolation \cite{mencik2024improved} was implemented as a reference, for which a two-dimensional scheme with 2nd order polynomials was employed. As the CB matrices are only interpolative within each region, the Gaussian points used as support points were located around the nominal model parameter space, with a perturbation magnitude $P=0.125$, similar to the implementation presented in Section~\ref{S:5.2}. A total of $3\times3=9$ support points were taken, and are presented in the model parameter space in Figure~\ref{F24}. 

\begin{figure}[ht]
    \centering
    \includegraphics[width=0.8\linewidth]{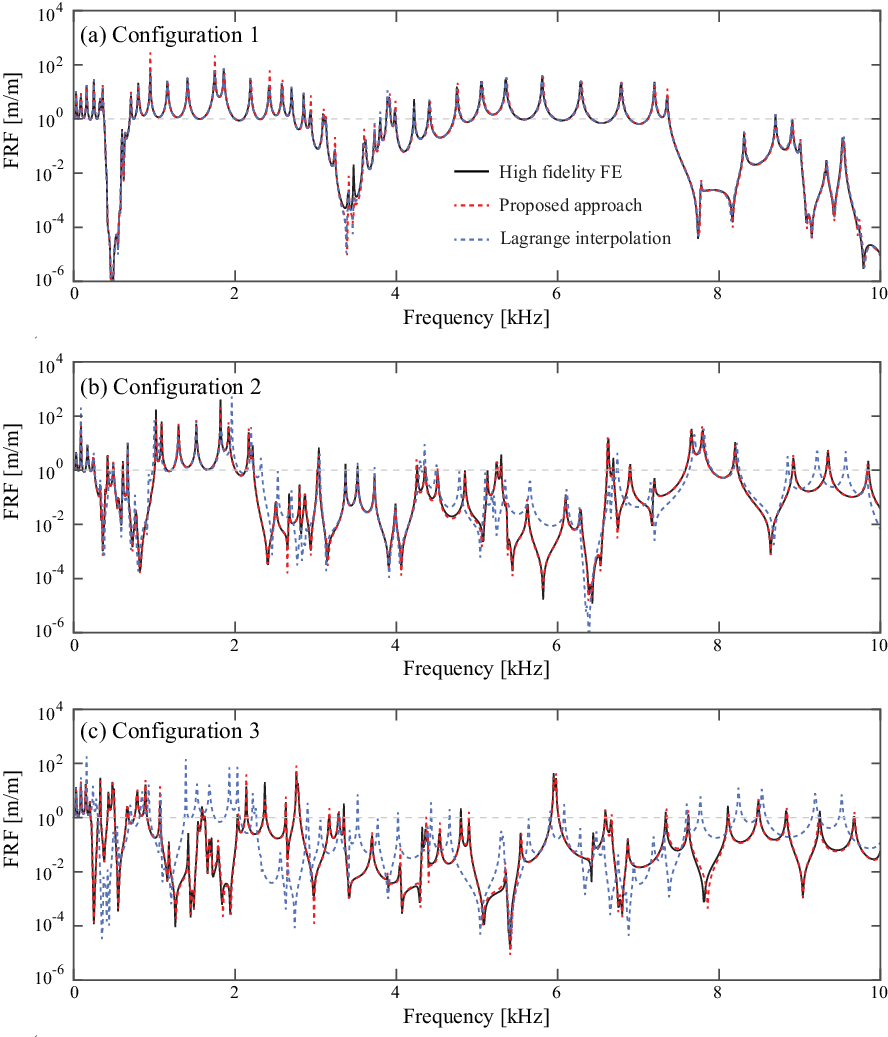}
    \caption{FRF of a structure composed of 10 substructures with varying model parameters subjected to perturbations around the nominal $\bfth^o$ of (a) 10\%, (b) 30\%, and (c) 50\%.}
    \label{F25}
\end{figure}

Finally, the FRF of each configuration is determined by using:(1) the proposed approach, (2) the Lagrange polynomials scheme, and (3) a high-fidelity FEM. The results for the predicted FRF are presented in Figure~\ref{F25}. Figure~\ref{F25}(a) depicts the results for Configuration 1, where the FRF estimation using both the proposed approach and Lagrange interpolation closely matches the high fidelity results. Figure~\ref{F25}(b) corresponds to Configuration 2, where some discrepancies in the identified FRF begin to appear for the Lagrange Polynomials strategy, while the proposed approach retains a close matching with the high fidelity results. These discrepancies can be attributed to two factors. First, there are substructures outside the supported design space, i.e., there are substructures outside the dashed box presented in Figure~\ref{F24}, which corresponds to extrapolations. Second, substructures identified as 3, 6, and 8 belong to regions different than $\bfTh^{o2}$; thus, the interpolation model trained in $\bfTh^{o2}$ cannot be used to predict matrices in other regions. Finally, Figure~\ref{F25}(c) presents the FRF for Configuration 3, where the discrepancies between high fidelity results and the Lagrange Polynomials further intensify. Nonetheless, for all three perturbation sizes, the proposed approach was capable of matching the FE results, highlighting the efficacy of the framework.

\newpage

\section{Conclusions}
\label{S:7}

This work introduces a significant advancement in surrogate models for predicting the frequency response of dynamic mechanical metamaterials, specifically addressing challenges posed by large parametric perturbations. While previous Finite Element (FE)-based matrix interpolation techniques, like Mencik's method \cite{mencik2021model, mencik2024improved}, efficiently allow substructure-level geometrical variations without remeshing, their effectiveness is limited by the inherent restrictions of the common modal projection technique. This limitation arises because the projection of Craig-Bampton (CB) matrices onto a common modal space inherently restricts the allowable perturbation size of model parameters, leading to ill-conditioning due to phenomena such as mode crossing. However, this prior work lacked to show explicit strategies to identify the usable region for valid interpolation or to extend the method beyond these boundaries. 

To overcome these restrictions, the present work makes three primary contributions. First, it provides structural dynamic insight into the limitations of the common modal projection, demonstrating that ill-conditioning can be controlled by strategically selecting the reference point $\bfth^o$. Second, it proposes an efficient, sampling-based procedure to identify the non-regular boundaries of the usable region in the model parameter space $\bfTh^o$ where the surrogate model can be reliably deployed. This procedure involves a multistage sampling approach combined with a Support Vector Machine (SVM) for classification to delineate the well-conditioned space. Within this identified region, Principal Component Analysis (PCA) is applied for dimensionality reduction of the output data (CB matrices) before training a Kriging interpolation model to predict these reduced matrices efficiently. This integrated approach offers an efficient workflow by reusing samples generated during boundary detection. Third, the work enhances the surrogate model to accommodate larger model parameter perturbations by proposing a multi-region framework. This procedure selects different reference sets of model parameters, enabling the identification and mapping of multiple usable subregions $\{\bfTh^{o1},\bfTh^{o2}...,\bfTh^{om}\}$ within the parameter space. For each region, a specific PCA-Kriging model is trained, and a multi-class SVM is employed to direct any new sample to the correct interpolation region. This innovative framework significantly extends the applicability of interpolation-based surrogate models for metamaterial design.

The efficacy of this proposed framework is thoroughly validated through two illustrative examples. The first example, involving a unit cell with a square plate and circular core, confirms the approach for a single well-conditioned projection region, showing excellent agreement with high-fidelity Finite Element Method (FEM) and Lagrange interpolation methods. The second example, using a beam-like structure with vibration attenuation bands, demonstrates the true advantage of the multi-region approach. In this more complex scenario, where predictions from traditional Lagrange interpolation deviated significantly with increasing perturbations, the proposed method maintained high accuracy across different perturbation levels, confirming its superiority for designing metamaterials with complex and varied geometries and enabling wider exploration of the design space.

In general, the proposed method offers significant advantages in identifying usable parameter spaces and accommodating larger perturbations by using a multi-region framework; one of its limitations is the increased requirement for support points/samples, especially for training the Kriging interpolation component to ensure accuracy. Computing these samples, which involves Finite Element Method (FEM) analysis for each substructure, is the most computationally expensive step. However, this initial investment offers substantial long-term benefits. Specifically, its ability to reliably explore a broader and often non-regular parameter space, even under large parametric perturbations, makes it exceptionally well-suited for applications requiring extensive recurrent analyses, such as Monte Carlo simulations for uncertainty quantification or iterative shape/topological optimization. In these demanding contexts, the increased accuracy and robustness across a wide design space can significantly reduce overall computational burden by minimizing the need for costly high-fidelity simulations during the optimization or uncertainty analysis loops.

\section*{Acknowledgments}

We acknowledge Dr. Mencik for providing the necessary information to replicate the numerical results extracted from \cite{mencik2024improved} and included in Section~\ref{S:5}. The research is funded by Dr. Ruiz's startup package at the University of Michigan-Dearborn.   

\bibliographystyle{model1-num-names}
\bibliography{references}
\end{document}